# On-chip manipulation of magnetic nanoparticles through domain walls conduits

P. Vavassori[1], M. Gobbi[4], M. Donolato[4], V. Metlushko[2], B. Ilic[3], , M. Cantoni[4], D. Petti, S. Brivio and R. Bertacco[4]

[1]CIC nanoGUNE Consolider, 20018 Donostia-San Sebastian, Spain and CNR-INFM S3, CNISM and Dipartimento di Fisica, Università di Ferrara, 44100 Ferrara, Italy

[2]Department of Electrical and Computer Engineering, University of Illinois at Chicago, Chicago, Illinois 60607

[3]Cornell Nanofabrication Facility, Cornell University, Ithaca, New York 14853

[4]LNESS – Dipartimento di Fisica Politecnico di Milano, Via Anzani 42, 22100 Como (Italy)

**Abstract**

The manipulation of geometrically constrained magnetic domain walls (DWs) in nanoscale magnetic strips has attracted much interest recently, with proposals for prospective memory [1] and logic devices [2].

Here we propose to use the high controllability of the motion of geometrically constrained DWs for the manipulation of individual nanoparticles on a chip with an active control of position at the nanometer scale. The proposed method exploits the fact that magnetic nanoparticles in solution can be captured by a DW, whose position can be manipulated with nanometric accuracy in a specifically designed magnetic nanowire structure. We show that the high control over DW nucleation, displacement, and annihilation processes in such structures can be used to capture, transport and release magnetic nanoparticles. As magnetic particles with functionalized surfaces are commonly used as molecule carriers or labels, the accurate control over the handling of the single magnetic nanoparticle is crucial for several applications including single molecule manipulation [3], separation [4], cells manipulation [5] and biomagnetic sensing [6].

In perspective, the easy integration on chip of nano-transport lines and sensors of domain walls and particles [7] allows for the realization of programmable circuits for molecular manipulation, analysis and synthesis, with continuous remote control of process.

Magnetic domain walls (DWs), which constitute the boundary between domains in ferromagnetic samples, have been intensively studied in the past, particularly in continuous thin films. The most prominent examples are the Bloch and Néel wall types [8]. In nanoscale confined stuctures, novel domain wall types emerge, when the wall spin structure starts to be dominated by the geometry rather than by the intrinsic material properties. For example, in nanoscale ferromagnetic wires (nanowires), shape anisotropy restricts the magnetization to lie parallel to the wire axis. In such a nanowire, a DW is a mobile interface, which separates regions of oppositely aligned magnetization. Each magnetic domain, has a head (positive or north pole) and a tail (negative or south pole). The resulting DWs are therefore either head-to-head (HH) or tail-to-tail (TT), and successive DWs along the nanowire alternate between HH and TT configurations. The motion of such geometrically confined DWs in nanoscale magnetic strips has become recently the focus of wide-spread theoretical [9,10] and experimental [1, 11-14] research for a number of reasons. Due to the geometrical confinement, the spin structure of a DW can be controlled via the lateral dimensions and film thickness of the nanowire, while the DW size is in the tens of nm range. Such DWs behave as quasi-particles, which can be precisely manipulated for instance by playing with external fields, spin-polarized currents and geometry. In particular variations in the geometry such as constrictions, protrusions or corners, determine a potential landscape for the DW, that can be engineered in order to obtain a finely controlled manipulation of the DW displacement. These aspects are also related to the potential for recently demonstrated technological application schemes that include magnetic random access memory [1] and logic devices [2].

Submicrometer planar strips (planar nanowires) made of a soft magnetic material such as Permalloy (Py, $Ni_{80}Fe_{20}$) have been shown to form excellent conduits for DWs that can be nucleated and manipulated in a high controllable way [1, 11-14]. In addition it has been shown that, under the action of an externally applied magnetic field, DWs can be propagated through complex 2D and 3D networks of nanowires that contain bifurcation and intersection points [1].

In this communication, another potential application that exploits the manipulation of geometrically constrained magnetic DW in nanoscale magnetic strips is presented. In particular we show how patterned planar magnetic nano-wires can be used to implement devices for the programmable capture, transport over large distances, accumulation, and release of nanometric magnetic particles in liquid environments, with active control in position (at the nm scale) and time.

In the last years other techniques have been developed both for the manipulation and transport of massive particle population or of a single particle, e.g., microfabricated current carrying wires [15], micromagnets [16], and magnetic tweezers [17]. More recently, arrays of magnetic elements patterned in thin film were also proposed for the transport of single magnetic particles exploiting the stray field generated by the magnetization of the elements and the manipulation of each element magnetization using an external magnetic field [18, 19]. In addition, it has been shown that a domain wall (DW) in a thin film can be used to manipulate magnetic micro-particles on a solid or fluid interface [20, 21].However, none of these techniques achieved the true single particle manipulation ability at the nanoscale provided by the approach described here.

This feature is particularly relevant because magnetic nanoparticles can be employed as labels or carriers of molecules. The manipulation of magnetic nano-particles then allows for precise molecular manipulation, a hot topic in chemistry, biology and medicine. In particular, working with small concentrations of the molecules of interest and with sufficiently small particles make it possible to attach a single molecule to the magnetic particle, opening the way to single molecule biophysics, viz, the study of biomolecular interactions at the level of individual molecules [3].

Besides the controlled and precise manipulation of DWs, the other basic ingredient of the application described here is the highly inhomogeneous magnetic stray field, of up to several kOe, generated by a DW. This stray field is spatially localized on the nanometer scale due to the DW's very confined geometric structure. Figure 1a shows a schematic of the spin configuration of a HH DW in a thin Py nanowire 20 nm-thick and 100 nm wide as calculated using the OOMMF public simulation platform [22]. In particular the DW structure shown in Fig. 1a refers to a 'transverse' HH DW wall as the magnetization at the center of the wall is transverse to the strip axis [9]. The gradient of the magnetic stray field generated by a DW in a nanowire results in an attractive force on any magnetic nanoparticle that is moving in proximity of the DW location. The magnetic nanoparticle can be either ferromagnetic or superparamagnetic: in the first case the strong DW field gradient orients and attracts the magnetic dipole moment of the particle; in the second case the DW field induces a magnetic moment in the particle that is then attracted by the field gradient towards the DW. Figure 1b shows a color intensity plot of the modulus of the attractive force together with a sketch of the potential energy surface experienced by a point magnetic dipole with a unit magnetic moment placed on a plane at 100 nm from the upper surface of the DW conduit structure. The plots have been calculated by computing, with OOMMF, the magnetic field **H** created in the surrounding space by the HH DW and using the following vector expression for the force: **F** = $\mu_0$(**μ•∇**) **H,** where **μ** is a unit magnetic moment (μ = 1 Am$^2$) placed in a plane at 100 nm from the nanowire surface. As an example, a typical value of μ of commercially available superparamagnetic nanoparticles is in the order of $10^{-16}$ - $10^{-15}$ Am$^2$, in which case the intensity of the attractive force can be estimated to be in the 10 – 100 pN range in a plane 100 nm above the conduit. The plots show that the gradient of the stray field emanating from the DW produces a self-focusing action that can trap a nano-particle flowing in the vicinity of the structure.

The use of the DW stray field for trapping nano- and micron-size superparamegnetic particles has been demonstrated experimentally in the case of Py nanorings [7] and in thin garnet films [15, 16]. In the latter case, it was also shown that a DW in a garnet thin film can be used to drag and manipulate magnetic colloidal particles at a solid - fluid interface, however with a limited control over the number of particles, the distance and the direction [20, 21].

The much higher control on the nucleation, displacement, and annihilation of DWs via external fields achievable in planar nanowires made from a soft magnetic material, induced us to explore the use of such structures as DW conduits to achieve a true nanometric manipulation of individual magnetic nano-particles in wet environments. Additionally, the use of planar nanowires as DW conduits is particularly interesting since it may, in the future, make it possible to move DWs via spin polarized current pulses [23].

In order to demonstrate the viability of the method, we first used square rings of Py, as the DWs motion in these structures is particularly simple and has been extensively investigated in previous articles [24]. They are 30 nm thick Py square rings with outside size of 6 μm and width of each segment of 200 nm (see Fig. 2b), e-beam lithographically patterned on top of a $SiO_2$/Si substrate and capped with a 50 nm thick $SiO_2$ protecting layer.

Two DWs, one HH and the other TT, are initially placed in the left upper and right bottom corners by applying an external field $H_0$ along the diagonal (left image in fig. 2a. Subsequently the DWs are displaced back and forth along the upper and lower segments of the ring by applying a horizontal field H, as shown by micromagnetic simulations and magnetic force microscopy (MFM) images in Fig. 2a. An array of structures, prepared in the initial state with the DWs placed in the upper left and bottom right corners, are then covered with a solution of NH4-OH (pH = 8) containing magnetic beads (nanomag®-D, 500 nm diameter) with a concentration of $10^6$ particles/μl, till some of the beads are captured by the DWs. The diameter of the particles was chosen to enable us to monitor, in real time, the beads displacement using optical microscopy. The left-hand side of the optical microscopy image in Fig. 2b shows the initial state with a cluster of magnetic nanoparticles captured by the HH DW in the upper left corner of two rings. When the horizontal external field H moves the DW from one corner to the other the particles also move to produce the configuration shown in the right-hand side of Fig. 2c, demonstrating that the magnetic nanoparticles follow the DW motion. While the time scale of DW motion in Py wires is very short, of the order of a ns over a length of a few microns [10, 11, 13, 14], in our experiment the displacement of the magnetic nanoparticles is much slower. With the time resolution of the camera we employed, we can estimate that the nanoparticle displacement takes place in a few hundreds of milliseconds. In this sense the nanoparticles do not strictly move *with* the DW but rather diffuse towards the potential energy minimum generated at the new position occupied by the DW after the application of H.

We repeated the experiment using Py square rings having the same thickness and width but different outside sizes between 1 and 6 μm: we found that the DW induced displacement of the nanoparticles is deterministic for the array of rings with outside size up to 2 μm. This upper limit is not intrinsic since it clearly depends on the ring material, thickness and width, the thickness and chemical properties of the capping layer, and the diameter and magnetic moment of the nanoparticle.

The DW displacement principle described for the square rings can be implemented in DW conduits made using nano-wires structures in order to displace magnetic nanoparticles over, in principle unlimited, larger distances. The zig-zag wire structure shown in the SEM image displayed in Fig. 3a, which implements a controllable magnetic DW step motor, is such an example. As for the square rings, the structure is a 30 nm thick Py wire e-beam lithographically patterned on top of a $SiO_2$/Si substrate and capped with a $SiO_2$ protecting layer 50 nm thick. The width of the wire is 200 nm while the width of the injection pad is 600 nm. Fig. 3b shows a sequence of magnetic force microscopy images and micromagnetic configurations of the injection and propagation of a domain wall under the action of external magnetic field pulses directed as sketched in Figure. The structure is prepared in a fully saturated initial state [configuration 1) in Fig. 3b] by applying a saturating field $H_0$ of 500 Oe as indicated in the Figure (note: the dark and bright portions on the left and right of the injection pad are not DWs, but they are only due to the stray field at the ends of the pad itself). Subsequently the nucleation pad is used for the nucleation of a reversed domain by applying a magnetic field $H_i$ of 100 Oe as sketched in Fig. 3b. A reversed domain nucleates in the pad at lower fields than elsewhere because the nucleation energy is lower due to its larger width and hence lower shape anisotropy than for the wire. The same magnetic field $H_i$ causes a HH DW to propagate to the first bend in the zig-zag conduit [configuration 2) in Fig. 3b]. The $H_i$ field is then removed and a sequence of fields $H_{UP}$ and $H_{DW}$ of 150 Oe, parallel to the wires branches are applied in order to displace the DW along each rectilinear segment and, thus, throughout the entire conduit [sequence of configurations 3) – 5) in Fig. 3b]. In order to employ this structure for manipulation of magnetic particles it is then prepared in the initial state with a HH DW placed in the first corner of the zig-zag conduit [configuration 3) in Fig. 3b] and the same solution, containing magnetic beads used for the previous experiment with the square rings, is dispensed on top of the structure, till one of the beads is captured by the DW. Figure 4 shows a sequence of frames of an optical [video1](video1) showing, in real time, the transport of a single magnetic particle along the conduit obtained applying a sequence of $H_{UP}$ and $H_{DW}$ fields as sketched in the Figure. The direction of motion can be reversed at any time by reversing the direction of $H_{UP}$ and $H_{DW}$ in the sequence. In addition, it is worth noticing that the corners are stable position for the DW so that a magnetic nanoparticle can, if required, be held in a selected position indefinitely.

In the two cases considered so far, the actual displacement between the stable DW positions occurs through a DW motion during which neither the DW structure nor the DW speed can be completely controlled. In this sense the motion takes place step by step, each step corresponding to the distance between neighboring pinning sites for DWs, and a continuous control of the motion is not allowed. Furthermore the very rapid DW displacement as well as the changes in the DW structure during its motion, e.g. from transverse-like to vortex -like [9, 10] that produces smaller stray fields, may result in the decoupling of the magnetic nanoparticle from the DW. These drawbacks can be avoided by using curved structures such as those shown in the next example. The employment of these circular rings allows us to induce a continuous and well-controlled DW displacement, but the same concept can be applied to curved portions of more complex nanowires. As in the case of the square ring, two DWs, one HH and the other TT, are spontaneously generated in a circular ring by applying a strong (saturating) magnetic field $H_0$. As shown by the OOMMF micromagnetic simulations in the top panel of Fig. 5 [configuration 1)] the DWs form upon removal of the field and lie along the direction that the field was applied. As shown in the top panel of Fig. 5, once created, the DWs can be moved around the circumference by the application of a smaller field H [configurations 2) an 3)]. The required magnitude is determined by the ring radius and the local DW pinning sites due to edge irregularities and material inhomogeneities as discussed below. By rotating the field both the DWs are displaced with the same angular speed as that of the rotating field, thus achieving smooth and fully controllable DW motion. In addition the constant presence of a field along the DW prevents any change in the transverse structure of the DW during its motion. Also in the present case the attractive force emanating from the DW causes the magnetic particles to follow the DW as shown by the sequence of optical microscopy images in the bottom panel of Fig. 5. The images show a Py circular rings of 10 μm diameter (thickness and width of 30 and 200 nm, respectively) where two clusters of magnetic nanoparticles, captured by the HH and TT DWs, are displaced by a rotating field H of 300 Oe. The continuous motion of the particles can be better appreciated from the [video2](video2) which shows two adjacent rings with only one cluster of particles each, one at a HH DW and the other at a TT DW. In all cases the particle follows the perimeter of the ring with its motion synchronous to that of the DW. Considering that the DW width can be in the range 10-100 nm depending on the magnetic material and the geometric confinement, we can conclude that a curved geometry allows for the control of the displacement and positioning of a magnetic nanoparticle with similar precision. This a crucial achievement in view of application to nano-manipulation of particles and molecules for fundamental investigations and synthesis.

The release of the magnetic particle requires the annihilation of the DW coupled to the magnetic nanoparticle. For zig-zag chains this can be achieved simply by tapering the end of the conduit thus producing weaker stray fields compared to a DW, as shown in Fig. 3b. Another solution that can be envisaged is to create intersections where a HH DW can merge with a TT DW causing the annihilation of the two DWs.

The concept presented here can be applied to more complex DW conduit network structures that exploit the possibility of duplicating a domain wall using a fan-out junction, making two split-up copies of the injected domain wall propagate in the two separate branches of the structure or the property that a domain wall traveling along a conduit can cross another conduit without being substantially modified. In this way multi-channel devices involving the motion of multiple DWs can be designed. The fabrication of conduits perpendicular to the plane would also allow for the creation of true 3D networks opening the way to devices with different layers of networks. Moreover, DW conduits structures can be devised to enable the simultaneous motion of multiple DWs to move several magnetic particles along the same transport line.

Another peculiarity of our approach is that one or more sensors of DWs and magnetic particles described in a previous work [7] can be fully integrated in a DW conduit as it essentially consists in a portion of the DW conduit, e.g., a corner, flanked by conductive contacts for detecting electrically the presence of a DW thanks to the anisotropic magnetoresistance effect. As shown in Ref. 14, the presence of a magnetic particle coupled to a DW is detected via the induced change in the magnetic field required for the displacement of the DW. Such a sensor could then be placed in the middle of a conduit and act as a magnetic particle counter, allowing for a digital control of the beads flowing on the conduits. This capability is unique compared to other technologies and pave the way to the realization of networks of conduits with externally programmable functions and continuous control of the desired process.

In perspective, the approach described here may support and integrate single molecule manipulation on chip with a much lower degree of complexity compared to the tools developed so far, which are highly sophisticated, require accurate calibration and are not suitable to be implemented in miniaturized devices [3].

Further, actual single-molecule techniques are generally limited to study one molecule at time preventing the possibility to sort and manipulate different targets and probe information about the statistical distributions of biological systems, while our approach has the capability to control several molecules at the same time on different structures of the same chip. Such a control would open up new possibilities particularly suited for lab-on-chip applications as, e.g., handling small volumes or performing sophisticated sorting of different targets attached to different particles.

In summary we have shown that patterned nanowire magnetic structures form domain wall conduits that can be tailored to form transport lines for magnetic particles exploiting the highly controllable motion of domain walls in such conduits and the magnetic coupling between a domain wall and a particle.

The careful design of the domain wall conduit allows the implementation of the following basic functionalities:

- capturing the magnetic nanoparticle at a desired position by domain wall creation or injection;

- motion of the magnetic nanoparticles along the magnetic domain wall conduit, where the domain wall can propagate with active control in position (at the nm scale) and time.

- releasing the magnetic nanoparticle in the solution at a desired position by domain wall annihilation

- parallel manipulation of nanoparticles in arrays of devices lithographed on the same chip

These functionalities can be fruitfully exploited in many fields where manipulation, i.e. capturing, movement, accumulation, and delivery, of nanometric magnetic particles is required within a lab-on-chip approach. A particularly relevant example, is the application of this method in biotechnology, where it has the potential of improving a number of existing applications, such as the transportation and sorting of biological molecules on surfaces, paving the way towards a microchip-based platforms for high-throughput single molecule analysis and synthesis.

**Figure Captions**

**Fig. 1** Panel a) shows a schematic of the spins configuration of a head-to-head domain wall in a thin Py nanowire 20 nm-thick and 100 nm-wide as calculated using the OOMMF public simulation platform. Panel b) shows a color intensity plot of the modulus of the attractive force per unit magnetic moment together with a sketch the potential energy surface on a plane at 100 nm from the Py nanowire in Panel a).

**Fig. 2** Panel a): micromagnetic simulations and magnetic force microscopy images of the stable magnetization configurations in a 6 μm x 6 μm Py square ring structure (thickness 30 nm and width 200 nm). Head-to-head and tail-to-tail domain walls are imaged by magnetic force microscopy as dark and bright spots, respectively. Panel b): optical microscopy images showing a cluster of a few magnetic nanoparticle (diameter 500 nm) captured by the head-to-head domain wall in the upper left corner of two rings of an array and their subsequent displacement to the upper right corner after the application of an horizontal magnetic field H.

**Fig. 3** Panel a): scanning electron microscopy image of the zig-zag wire structure made of Py used to implement a controllable magnetic domain wall step motor (thickness 30 nm, width of the zig-zag 200 nm, width of the nucleation pad 600 nm). Panel b): sequence of magnetic force microscopy images and micromagnetic configurations showing the injection and propagation of a domain wall under the action of external magnetic fields $H_{UP}$ and $H_{DW}$ directed as sketched in Figure. The dark and bright portions on the left and right of the injection pad are not DWs, but they are only due to the stray field at the ends of the pad itself.

**Fig. 4** Sequence of optical microscopy images of the transport of a magnetic particle along the wire obtained applying a sequence of fields $H_{UP}$ and $H_{DW}$ as sketched in the Figure.

**Fig. 5** Top Panel : micromagnetic simulations of the nucleation and displacement of head-to-head and tail-to-tail domain walls in a circular ring obtained by applying a rotating field H. Bottom panel: optical microscopy images of the displacement of two magnetic particles, captured by the two domain walls in a 10 µm diameter Py ring (thickness 30 nm, width 200 nm) by applying a rotating field H of 300 Oe.

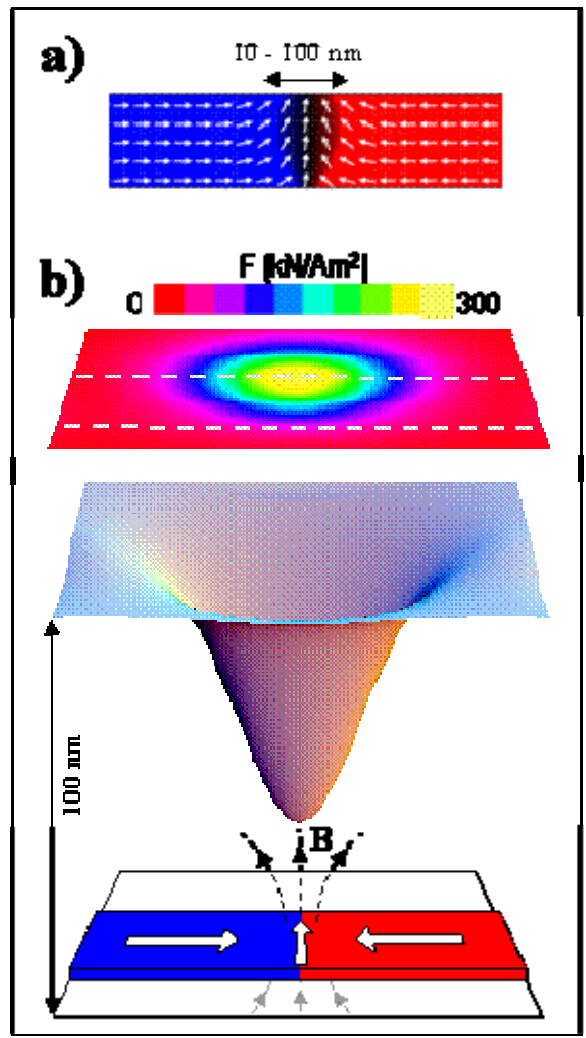

Fig 1

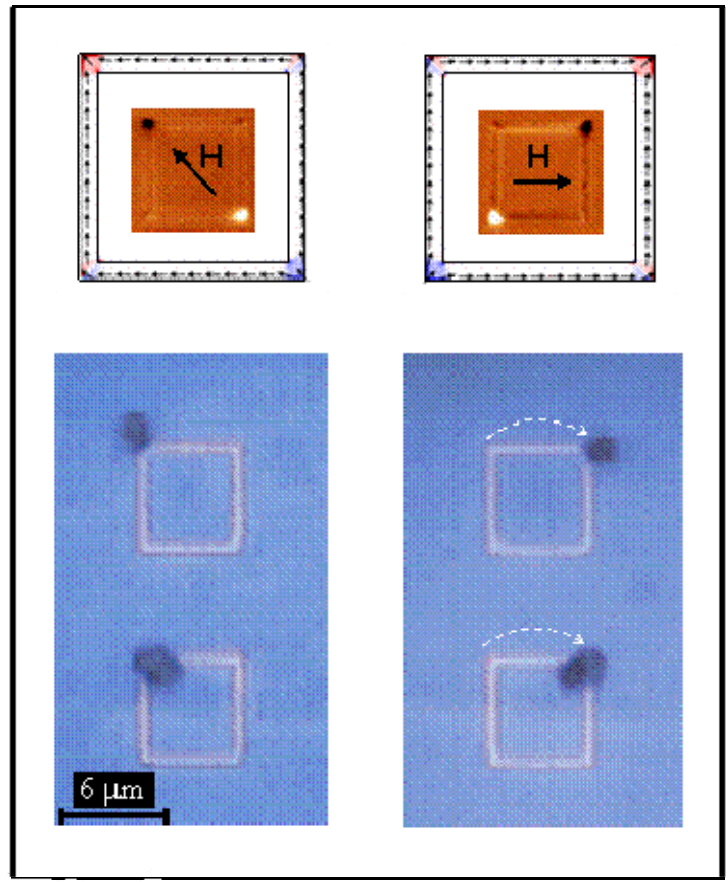

Fig 2

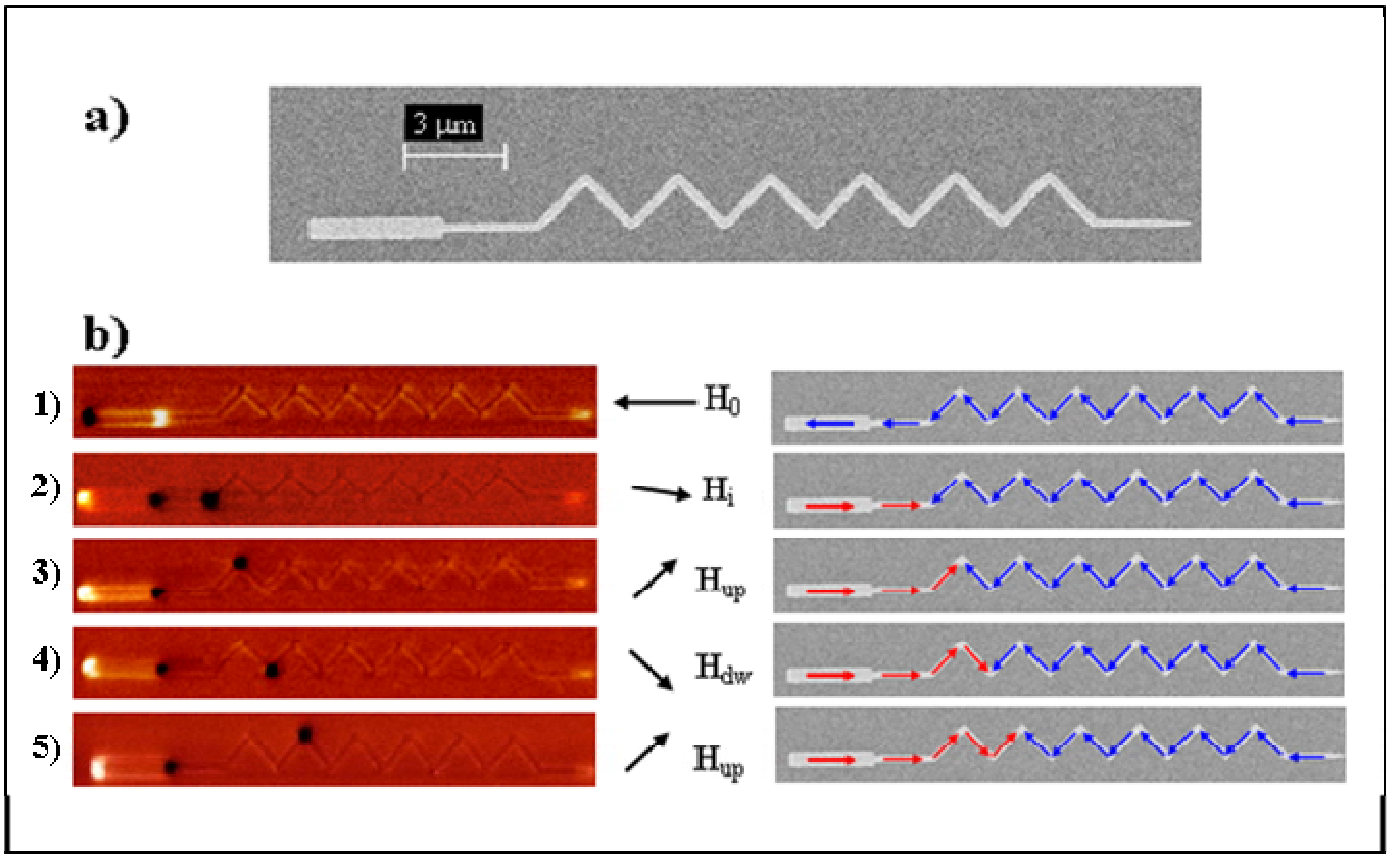

Fig 3

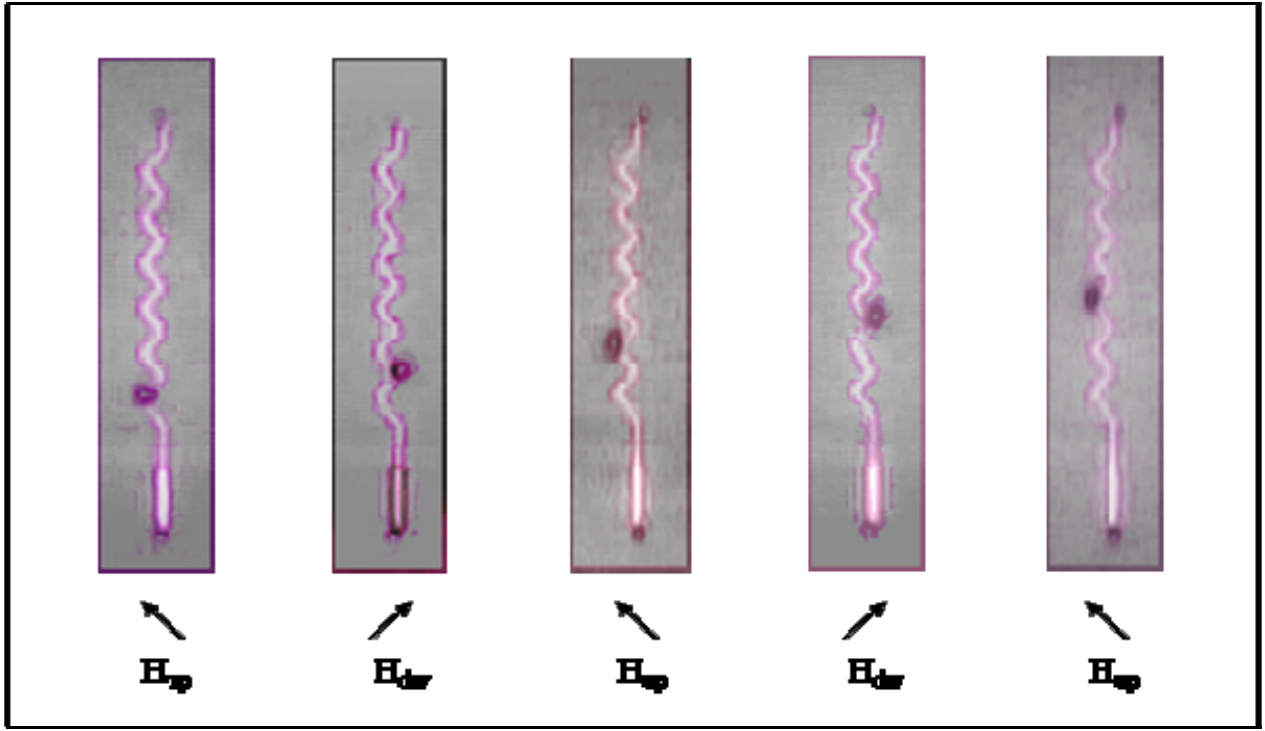

Fig 4

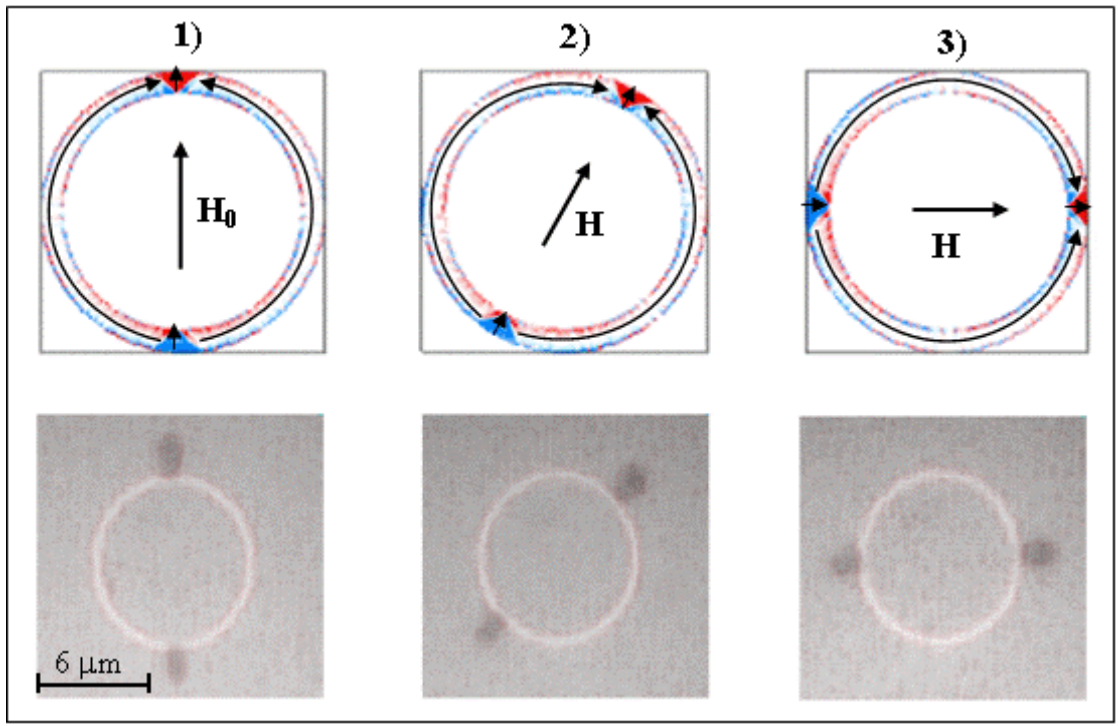

Fig 5